\documentclass[a4paper,fleqn,usenatbib]{mnras}

\usepackage{newtxtext,newtxmath}

\usepackage[T1]{fontenc}
\usepackage{ae,aecompl}

\usepackage{graphicx}    
\usepackage{amsmath}    
\usepackage{amssymb}    

\usepackage{todonotes}
\graphicspath{{figures/}}   

\newcommand{\Sun}[0]{\ensuremath{_{\odot}}}

\defcitealias{2017MNRAS.467..179G}{G17}

\title[Auriga GCS]{The globular cluster system of the Auriga simulations}

\author[T. L. R. Halbesma et al.]{\parbox[t]{\textwidth}{
    Timo L. R. Halbesma$^{1}$\thanks{E-mail: Halbesma@MPA-Garching.MPG.DE},
    Robert J. J. Grand$^{1}$,
    Facundo A. G\'{o}mez$^{2,3}$,
    Federico Marinacci$^{4}$,
    R\"{u}diger Pakmor$^{1}$,
    Wilma H. Trick$^{1}$,
    Philipp Busch$^{1}$,
    Simon D. M. White$^{1}$
} \vspace{10pt} \\
$^{1}$ Max-Planck-Institut f\"ur Astrophysik, Karl-Schwarzschild-Str. 1,
    85741 Garching, Germany \\
$^{2}$ Instituto de Investigaci\'{o}n Multidisciplinar en Ciencia yTecnolog\'{i}a,
    Universidad de La Serena, Ra\'{u}l Bitr\'{a}n 1305, La Serena, Chile \\
$^{3}$ Departamento de F\'{i}sica y Astronom\'{i}a, Universidad de La Serena, Av.
    Juan Cisternas 1200 N, La Serena, Chile \\
$^{4}$ Department of Physics \& Astronomy, University of Bologna, via Gobetti 93/2, 40129 Bologna, Italy \\
}

\date{Accepted XXX. Received YYY; in original form ZZZ}

\pubyear{2019}

\begin{document}
\label{firstpage}
\pagerange{\pageref{firstpage}--\pageref{lastpage}}
\maketitle

\begin{abstract}
We investigate whether the galaxy and star formation model used for the Auriga
simulations can produce a realistic globular cluster (GC) population. 
We compare statistics of GC 
candidate star particles in the Auriga haloes with catalogues of the Milky Way 
(MW) and Andromeda (M31) GC populations. We find that the Auriga simulations do 
produce sufficient stellar mass for GC candidates at radii and metallicities that 
are typical for the MW GC system (GCS). We also find varying mass-ratios of the
simulated GC candidates relative to the observed mass in the MW and M31 GC systems
for different bins of galactocentric radius-metallicity ($r_{\text{gal}}$-[Fe/H]).
Overall, the Auriga 
simulations produce GC candidates with higher metallicities than the MW and M31 
GCS and they are found at larger radii than observed. The Auriga simulations would
require bound cluster formation efficiencies higher than ten percent for the 
metal-poor GC candidates, and those within the Solar radius should experience 
negligible destruction rates to be consistent with observations.
GC candidates in the outer halo, on the other hand, should either have low formation
efficiencies, or experience high mass loss for the Auriga simulations to produce a
GCS that is consistent with that of the MW or M31. Finally, the scatter in the 
metallicity as well as in the radial distribution between different Auriga runs 
is considerably smaller than the differences between that of the MW and M31 GCSs. 
The Auriga model is unlikely to give rise to a GCS that can be consistent with
both galaxies.
\end{abstract}




\section{Introduction}
Globular star clusters (GC)s are old, bright, and ubiquitous. Various properties of
GC systems (GCS) show correlations with their host galaxies. GCSs are believed to
retain information about the galactic (gas) conditions at their times of formation.
Thus they could offer a unique insight into the (chemodynamical) evolution of their 
parent galaxies, if the formation and evolution of GCs- and GC systems themselves 
is adequately understood. However, despite decades of research (e.g.~\citealt{
1991ARA&A..29..543H, Harris2001, 2004Natur.427...31W, 2006ARA&A..44..193B, 
2012A&ARv..20...50G, 2014CQGra..31x4006K, 2018RSPSA.47470616F}), consensus on an 
exhaustive picture of the formation of GCs is yet to be reached. Most galaxies host
a GCS with a bimodal colour distribution that may reflect a bimodal metallicity 
distribution which is argued to arise from two distinct formation channels 
\citep{2006ARA&A..44..193B}. GCs are typically referred to as `blue' or `red' \citep[e.g.][]{
1985ApJ...293..424Z,1999AJ....118.1526G,2001AJ....121.2974L} where the former class
is more metal-poor, more radially extended, and showing little to no rotation whereas
the latter is more metal-rich, radially concentrated, and rotating with the galaxy.

The literature offers a wealth of formation scenarios, many of which focus on the
distinct characteristics of the blue or red subpopulation. \citet{1968ApJ...154..891P}
and \citet{1984ApJ...277..470P} argue that GCs form as the earliest bound
structures in the Universe (i.e. prior to formation of the main galaxy), 
noting that the Jeans length and mass shortly after recombination is consistent
with typical GC masses and sizes. \citet{2005MNRAS.364..367D} and \citet{
2009ApJ...706L.192B} argue that the blue, metal-poor GCs ([Fe/H]$ < -1$) form in 
radially biased dark matter (DM) halos at high redshift. Other recent hypotheses 
of GC formation prior to collapse of the proto-galaxy include formation in colliding 
supershells \citep{2017Ap&SS.362..183R}, in supersonically induced gaseous objects
\citep{2019arXiv190408941C}, or in high-speed collisions of dark matter subhaloes
\citep{2019arXiv190508951M}. \citet{2017MNRAS.472.3120B} suggests
that (the blue subpopulation of) GCs could form in high density regions along
the cosmic filament before or during collapse.

Other models also date GCs formation during formation of the proto-galaxy itself,
for example as a result of thermal instabilities in hot gas-rich haloes 
(\citealt{1985ApJ...298...18F}, also see the discussions in \citealt{
1990ApJ...363..488K}). Alternative formation triggers are also explored, such as 
(other causes of) shock compression, or cloud-cloud colissions \citep[e.g.][]{
1980glcl.conf..301G, 1992ApJ...400..265M, 1994ApJ...429..177H, 1995ApJ...442..618V,
1996ASPC...92..241L, 2001ApJ...560..592C}. 

Yet another hypothesis is that star cluster formation traces periods of high-intensity
star formation, which can be triggered by major gas-rich (spiral) galaxy mergers 
\citep{1987nngp.proc...18S, 1992ApJ...384...50A}. Such a scenario is naturally 
expected within the framework of hierarchical assembly and predicts formation 
of young clusters in interacting and merging galaxies, which has been observed 
and are found to show remarkable similarities with GCs in the Milky Way (MW) \citep[e.g.][]{
1995AJ....109..960W, 1996AJ....112..416H, 1999AJ....118..752Z, 1999AJ....118.1551W}.
Moreover, modelling efforts of this framework yield GC (sub)populations consistent 
with various observables \citep[e.g.][]{2010ApJ...718.1266M, 2018MNRAS.480.2343C}, 
and the recent numerical simulation of an isolated dwarf-dwarf merger executed at 
very high resolution (baryonic mass $m_b \sim$~4M\Sun; softening $\epsilon = 0.1$~pc) 
produces star clusters that could be GC progenitors \citep{
2019arXiv190509840L}.

As for the formation timeline, the scenarios (and flavours thereof) of \citet{
1985ApJ...298...18F} and \citet{1992ApJ...384...50A} are intertwined because
accretion and mergers continually occur during the hierarchical build-up of
galaxies. Various (other) hierarchical formation channels thus combine different
aspects of the aforementioned paradigms, such as GC formation in (small) galactic
disks before they are accreted onto an assembling galaxy \citep[e.g.][]{
2000ApJ...533..869C, 2002ApJ...567..853C, 2002MNRAS.333..383B, 2003egcs.conf..224G}. 
We refer to \citet{2001astro.ph..8034G} for a scoreboard of GC formation models 
compared to observations of the MW GCSs.

An idea that has recently been studied extensively is the hypothesis that the formation
mechanisms of young massive clusters \citep[YMCs, see][for a review]{2010ARA&A..48..431P} 
and GCs are the same. The proximity of YMCs in the local Universe allows to study these 
processes with a level of detail that cannot be achieved for GCs at high redshifts. In 
this framework, differences between the two classes of objects are caused by nearly a 
Hubble time of (dynamical) evolution \citep[e.g.][]{1987degc.book.....S}. This picture 
is based on observed similarities between YMCs and GCs
\citep[e.g.][]{1992AJ....103..691H,1999AJ....118.1551W} and strengthened by observations
of gravitationally lensed objects at high redshifts ($z = 2-6)$. These sources have 
properties reminiscent of (local) YMCs and may be GC progenitors at times of formation 
\citep{2017MNRAS.467.4304V,2017ApJ...843L..21J}. The modelling work by 
\citet{2011MNRAS.414.1339K,2012MNRAS.421.1927K,2015MNRAS.454.1658K} is now 
incorporated into cosmological zoom simulations and shows promising results
\citep{2018MNRAS.475.4309P,2019MNRAS.486.3134K}.

The (mass) resolution of cosmological zoom simulations has reached the mass range 
populated by GCs, and the gravitational force softening can be as low as several 
parsec. A number of groups can thus incorporate formation of (globular) star 
clusters into their high-resolution hydrodynamical simulations. \citet{2016ApJ...831..204R} 
ran parsec scale simulations of the high-redshift universe 
prior to reionization (the simulations stop at $z=9$), \citet{2017ApJ...834...69L} 
implement a new subgrid model for star (cluster) formation and run simulations 
that reach $z=3.3$, and the run of \citet{2017MNRAS.465.3622R} reaches $z=0.5$. 
\citet{2018MNRAS.474.4232K} find that mergers can push gas to high density that
quickly forms clustered stars that end up tightly bound by the end of the 
simulation. A somewhat different approach couples semi-analytical models to DM-only
simulations \citep{2010ApJ...718.1266M,2014ApJ...796...10L,2018MNRAS.480.2343C,
2019MNRAS.486..331C,2019arXiv190505199C}, also done in the work by 
\citet{2019MNRAS.482.4528E}.

In this work we use state of the art simulations that produce realistic spiral 
galaxies (MW analogues) at redshift zero for which several global properties 
are consistent with the observations. The question thus naturally arises whether 
the star formation histories of the simulations could give rise to a GCS similar to 
the MW and/or Andromeda (M31) GCSs. In particular, 
we use the Auriga simulations \citep[][hereafter G17]{2017MNRAS.467..179G}, 
further described in Section~\ref{sec:auriga}. We investigate whether the star 
formation model implemented produces metallicity, radial, and metallicity-radial 
distributions that are consistent with the MW and/or M31 GCSs, and whether the 
model produces enough stellar mass with the right properties. Moreover, the simulations
provide us with the history of the star particles, which allows us to investigate
the differences between these properties for sub-populations that have formed 
{\it in~situ} versus those that have been accreted. We focus on the metallicity
and radial distributions because observations of the red and blue GCs show
distinct differences between these properties.

The plan of the paper is as follows. We summarise the relevant characteristics 
of the Auriga simulations in Section~\ref{sec:auriga}, followed by a summary of 
the observations of the MW and M31 GCSs
in Section~\ref{sec:observations} that we compare our simulations to in
Section~\ref{sec:results}. We discuss our findings in Section~\ref{sec:discussion}
to come to our conclusions in Section~\ref{sec:conclusions}.

\section{The Auriga simulations}
\label{sec:auriga}
We use the Auriga simulations \citepalias{2017MNRAS.467..179G}, a suite of 
high-resolution cosmological zoom simulations of MW analogues. The MW-mass
selected initial conditions in the range $1-2 \times 10^{12}$~M$_{\odot}$
were selected from the \mbox{\textsc{eagle}} DM-only box, see 
\citetalias{2017MNRAS.467..179G} for details. The simulations are 
performed with the \textsc{arepo} code \citep{2010MNRAS.401..791S, 2016MNRAS.455.1134P} 
that solves the magnetohydrodynamical equations on a moving mesh. 

The galaxy formation model includes primordial and metal-line cooling with 
self-shielding corrections. Reionization is completed at redshift $6$ by a
time-varying spatially uniform UV background \citep{2009ApJ...703.1416F, 
2013MNRAS.436.3031V}. The interstellar medium (ISM) is described by an equation of 
state for a two-phase medium in pressure equilibrium \citep{2003MNRAS.339..289S}: 
cold clouds embedded in a hot ambient medium, whose mass and energy densities
are described by processes important for star formation, such as the condensation 
of cold clouds and the photoevaporation of clouds. Stars form stochastically in 
thermally unstable gas with a density threshold of $n = 0.13 \, \text{cm}^{-3}$,
and consecutive stellar evolution is accounted for. The channel for energetic
stellar feedback is from non-local type II supernovae (SN), which is modelled 
with wind particles (\citealt{2014MNRAS.437.1750M}; \citetalias{2017MNRAS.467..179G}).
Galactic winds enrich the ISM with metals from SNIa, SNII, and asymptotic giant 
branch stars \citep{2013MNRAS.436.3031V}. Supermassive
black holes grow via Bondi–Hoyle–Lyttleton accretion \citep{1944MNRAS.104..273B,
1952MNRAS.112..195B}, and feedback from the active galactic nuclei has a low-accretion
and high-accretion mode (radio and quasar) (\citealt{2005MNRAS.361..776S, 
2014MNRAS.437.1750M}; \citetalias{2017MNRAS.467..179G}). Finally, the simulations 
follow the evolution of a magnetic field of $10^{-14}$ (comoving)~G seeded at 
$z = 127$ \citep{2013MNRAS.432..176P, 2014ApJ...783L..20P}.

This model was tailored to the \textsc{arepo} code and calibrated to reproduce
key observables of galaxies, such as the history of the cosmic star formation rate
density, the stellar mass to halo mass relation, and galaxy luminosity functions.
The galaxy formation model produces realistic spiral galaxies at redshift zero
that match observations such as sizes, morphologies (\citetalias{2017MNRAS.467..179G}),
HI disc properties \citep{2017MNRAS.466.3859M}, and the diversity of accreted stellar
halo properties \citep{2019MNRAS.485.2589M}. Furthermore, the simulations are able to
reproduce major Galactic components, such as the chemical thin/thick disc dichotomy
\citep{2018MNRAS.474.3629G} and the recently discovered \textit{Gaia} Sausage 
\citep{2019MNRAS.484.4471F}. They are thus a large set of detailed simulations
similar enough to the Galaxy to make predictions for the GC population properties
of the MW and M31.

The Auriga suite has a fiducial baryonic mass resolution $m_b$~=~$5 \times 10^4$~M\Sun,
with gravitational softening of collisionless particles $\epsilon$~=~369~pc.
Simulations at this resolution are referred to as level 4 (L4). Selected
initial condition runs are acompanied by the lower (higher) resolution level L5 (L3).
The resolution for these levels is $m_b$~=~$4 \times 10^5$~M\Sun \, with 
$\epsilon$~=~738~pc, and $m_b$~=~$6 \times 10^3$~M\Sun \, with $\epsilon$~=~184~pc,
respectively. The mass resolution of the Auriga simulations is thus close $10^{5}$~M\Sun,
to the characteristic peak mass of the lognormal GC mass distribution 
\citep{1991ARA&A..29..543H}, although the gravitational softening is two orders 
of magnitudes larger than typical GC radii of several parsec. High-density gaseous 
regions are thus not expected to produce surviving stellar clumps with masses and 
radii consistent with GCs because such objects would numerically disperse, even 
in the highest-resolution runs. On the other hand, we can investigate (statistical) 
properties of age-selected GC candidates because each star particle represents a
single stellar population with a total mass that could be consistent with one GC. 
This means that their formation sites could be consistent with those of real-world GCs.

\section{Observational data}
\label{sec:observations}
We describe observations of the MW GCS in Section~\ref{sec:milkyway} for these clusters,
and of the M31 GCS in Section~\ref{sec:andromeda}. We discuss age-estimates
in Section~\ref{sec:age}, their distribution of galactocentric radius in 
Section~\ref{sec:Rgc}, and their total mass in 2D bins of metallicity-galactocentric
radius in Section~\ref{sec:observations_FeHRgc}. We focus on both metallicity and
galactocentric radius because these data are available in the observations and
in the Auriga simulations.

\subsection{Milky Way}
\label{sec:milkyway}
\citet[][2010 edition; hereafter H96e10]{1996AJ....112.1487H} provides a
catalogue\footnote{See \url{https://www.physics.mcmaster.ca/Fac_Harris/mwgc.dat}}
of the MW GCS that contains properties of 157 GCs. 
The authors initially estimated the number of GCs in the MW GCS to be 180~$\pm$~10,
thus, their catalogue to be ${\sim}$85\% complete. However, an additional 59 GC
candidates have been discovered more recently by various authors. The total number of GCs
in the MW might thus be 216 with recent estimates now anticipating an additional 
$30$~GCs yet to be discovered \citep[e.g.][and references therein]{2018ApJ...863L..38R}.
We still use data from the Harris catalogue, but caution that it may (only) be
50-70\% complete. Specifically, the relevant data fields that we use from H96e10
are the metallicity [Fe/H], the Galactic distance components $X$, $Y$, and $Z$ (in
kpc)\footnote{In a Sun-centered coordinate system: $X$ points toward Galactic
center, $Y$ in direction of Galactic rotation, and $Z$ toward the North Galactic
Pole. We calculate the galactocentric radius $r_{\text{gal}}=\sqrt{(X-r_\odot)^2
+ Y^2 + Z^2}$, assuming the solar radius $r_\odot=8$~kpc.}, and absolute
magnitude in the V~band, $M_V$. We use the V~band magnitude to calculate 
mass-estimates by assuming $M_{V,\odot}=4.83$ and a mass to light ratio 
$M/L_V = 1.7$~M$_{\odot}$/L$_{\odot}$, the mean for MW clusters \citep{2005ApJS..161..304M}. 
We supplement the catalogue with age-estimates from isochrone fits to stars near 
the main-sequence turnoff in 55 GCs \citep[][hereafter V13]{2013ApJ...775..134V}.


\subsection{Andromeda}
\label{sec:andromeda}
\citet[][hereafter C11]{2011AJ....141...61C} and \citet[][hereafter CR16]{2016ApJ...824...42C}
present a uniform set of spectroscopic observations of the inner 
$1.6^\circ~({\sim}21)$~kpc of M31 that is believed to be 94\% complete. In addition,
the outer stellar halo of M31 up to $r_{\text{proj}}\sim150$~kpc is observed in 
the Pan-Andromeda Archaeological Survey \citep[PAndAS, ][hereafter H14]{2014MNRAS.442.2165H,
2014MNRAS.442.2929V, 2019MNRAS.484.1756M}. In fact, 
the work of H14 is incorporated in the latest public release\footnote{Last 
revised 23 Sep 2015, see \url{https://www.cfa.harvard.edu/oir/eg/m31clusters/M31_Hectospec.html}}
of the C11 and CR16 data sets. Therefore we use the latest data set of CR16 
because it is the most recent aggregated dataset of M31's GCS that contains GCs 
in the inner region and in the outer halo. The relevant fields in the CR16 dataset 
that we use are the age, metallicity, and the mass-estimate\footnote{The authors 
assumed $M/L_V = 2$ independent of [Fe/H]}. We calculate the galactocentric radii 
from the observed positions RA and DEC, as further discussed in Section~\ref{sec:Rgc}.

\subsection{Age-estimates}
\label{sec:age}
The top panel of Figure~\ref{fig:MW-M31-age-Rgc} shows a histogram of the age-estimates
of the $55$ MW GCs in V13 and $88$ GCs in M31 for which age-estimates are available
in CR16. The mean age of the MW GCs in this data set is $11.9$~Gyr and the 
standard deviation is $0.8$~Gyr. Furthermore, only 1 of the 55 GC age-estimates is
below $10$~Gyr. The M31 GCS has a mean age of $11.0$~Gyr with a standard deviation 
of $2.2$~Gyr, and $24$ GCs have age-estimates below $10$~Gyr with a minimum age 
of $4.8$~Gyr. Based on these data, we find that the age distributions of the MW 
and M31 GCSs are not statistically consistent. The MW appears 
to host a GCS that is somewhat older than that of M31.
This is somewhat surprising given that M31 is generally considered to be earlier
type than the MW so that an older stellar population would naively be expected.
On the other hand, the MW could be an outlier as it may have formed and assembled 
most of its mass earlier than galaxies with a similar mass \citep[e.g.][]{2014ApJ...781L..31S,
2015A&A...578A..87S,2018MNRAS.477.5072M}.
In addition, we caution that both data sets are incomplete and the age 
measurements have large uncertainties (of $1-2$~Gyr). On the other hand, the 
magnitude of the uncertainty is insufficient to explain the low-age tail in M31.

\subsection{Radial distribution}
\label{sec:Rgc}
\begin{figure}
    \includegraphics[width=\columnwidth]{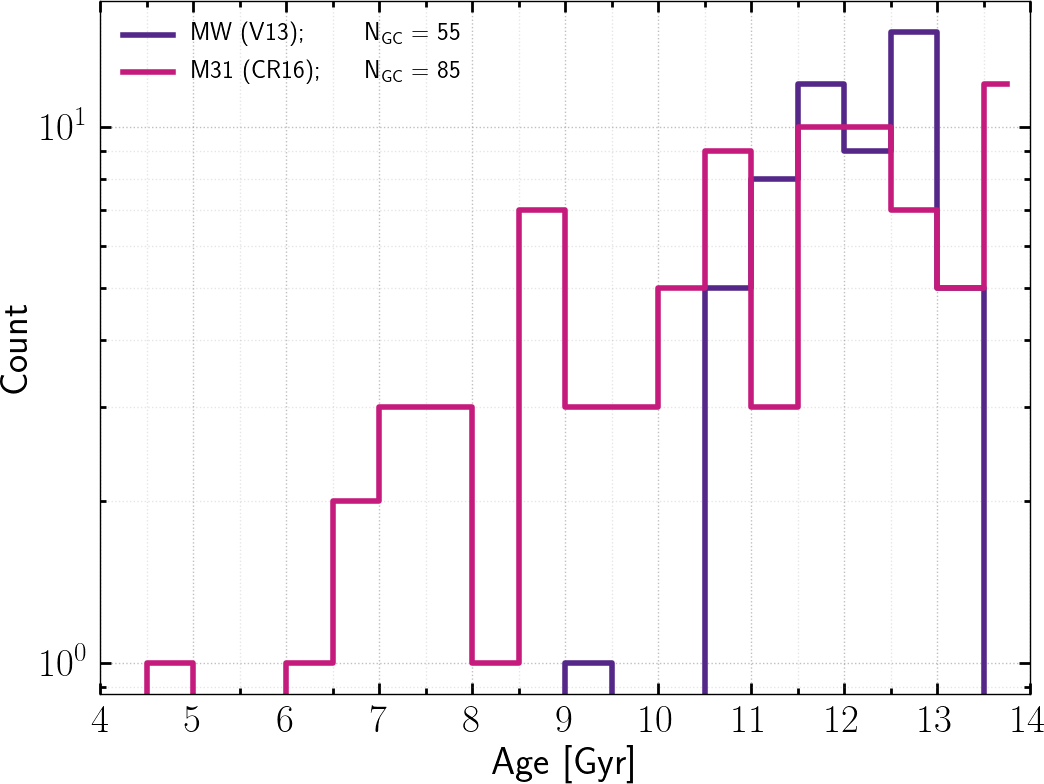}
    \includegraphics[width=\columnwidth]{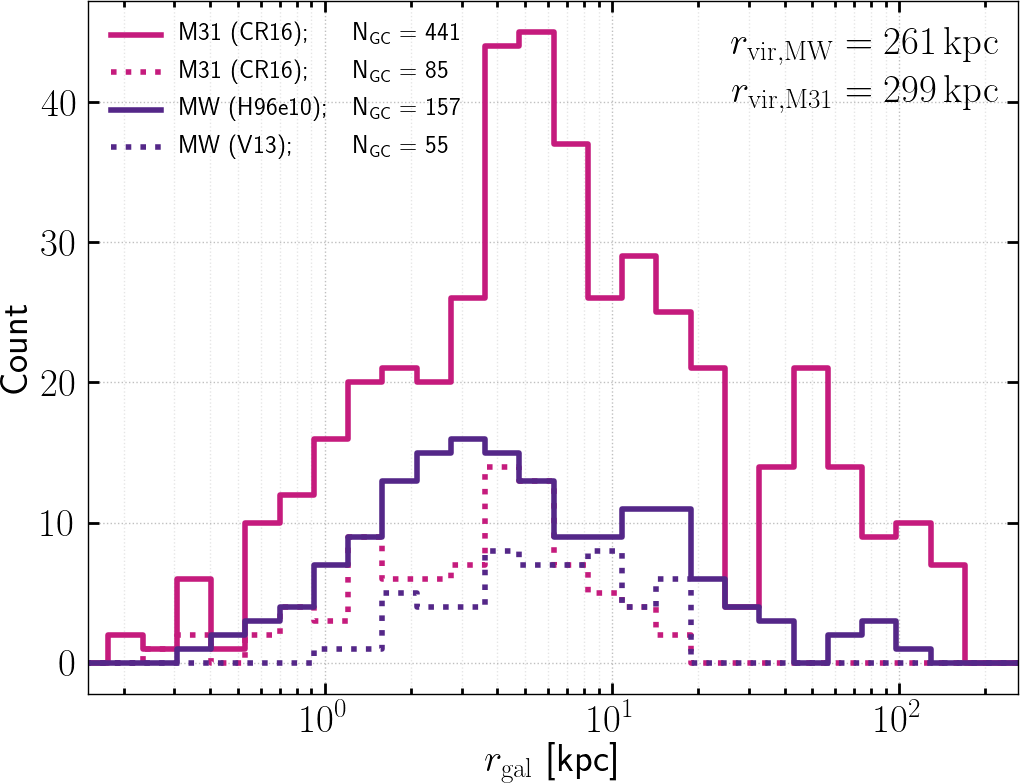}
    \caption{
        \emph{Top}: Age distribution of 55 GCs in the MW
        \citep[data from][]{2013ApJ...775..134V} and 88 GCs
        in M31 \citep[data from][]{2016ApJ...824...42C}.
        \emph{Bottom:} Distribution of galactocentric radius in the MW and M31.
        The dotted lines show the subset of data that also have age measurements
        (i.e. the same sample as in the top panel). We divide the radii of M31 by
        $r_{\text{vir,M31}} = 299$~kpc and multiply by $r_{\text{vir,MW}} = 261$~kpc 
        \citep{2017MNRAS.464.3825P} to compensate for the different intrinsic 
        sizes of both galaxies. 
        \label{fig:MW-M31-age-Rgc}
    }
\end{figure}

The bottom panel of Figure~\ref{fig:MW-M31-age-Rgc} shows the radial distribution
of the MW and M31 GCS. We divide the radii of M31 by its virial radius\footnote{
$r_{\text{vir}}$ is the radius in a spherical `top-hat' perturbation model at which
the average density reaches an overdensity of $\Delta_{\text{vir}}=357$ times the 
background density} $r_{\text{vir,M31}} = 299$~kpc and multiply by 
$r_{\text{vir,MW}} = 261$~kpc to account for the different intrinsic sizes of the two
galaxies, adopting the values and cosmology from \citet{2017MNRAS.464.3825P}.
Values for $r_{\text{gal}}$ are readily available in H96e10 (assuming $r_{\odot}=8.0$~kpc), 
but the galactocentric radius of GCs in M31 is not available in CR16. Therefore, we 
follow \citet[][section~4.1]{2019arXiv190111229W} 
to calculate the projected radius $r_{\text{proj}}$ from the observed positions,
adopting M31's central position from the NASA Extragalactic 
Database\footnote{\url{https://ned.ipac.caltech.edu/}} $(\alpha_0, \, \delta_0) =
(0^{\text{h}}42^{\text{m}}44.35^{\text{s}}, \, +41^{\circ}16'08.63")$
and the distance $D_{\text{M31}} = 780$~kpc \citep{2005MNRAS.356..979M,2012ApJ...758...11C}.
We estimate $r_{\text{gal}}$ as `average deprojected distance`
$r_{\text{gal}} = r_{\text{proj}} \times (4/\pi)$. 

The solid lines show the distributions using all available data (because the
sky coordinates are known for each GC), while the subset for which age-estimates
are available is indicated using dotted lines. The latter shows a narrower range
of radii than the full data set: it appears that few age-estimates are available 
for the innermost ($<1$~kpc) GCs, and none for those beyond roughly $20$~kpc
(the halo GCs). We compare the two distributions of the full data set (solid 
lines) and find that the MW has more GCs in the range $1-4$~kpc than M31
when accounting for the larger number of total GCs in M31. Interestingly, the two 
distributions show a similar trend for $r_{\text{gal}} > 4$~kpc and host a 
subpopulation of halo GCs. Our naive expectation would be that the spiral galaxies
in the Local Group host similar GC systems. However, based on a two-sample KS test
we find that the radial distributions of the GCS of the MW and M31 are not statistically 
consistent with being drawn from the same underlying distribution due to substantial 
differences at intermediate radii. 

\subsection{Total GC mass in metallicity-radial space} 
\label{sec:observations_FeHRgc}
\begin{figure}
    \includegraphics[width=\columnwidth]
        {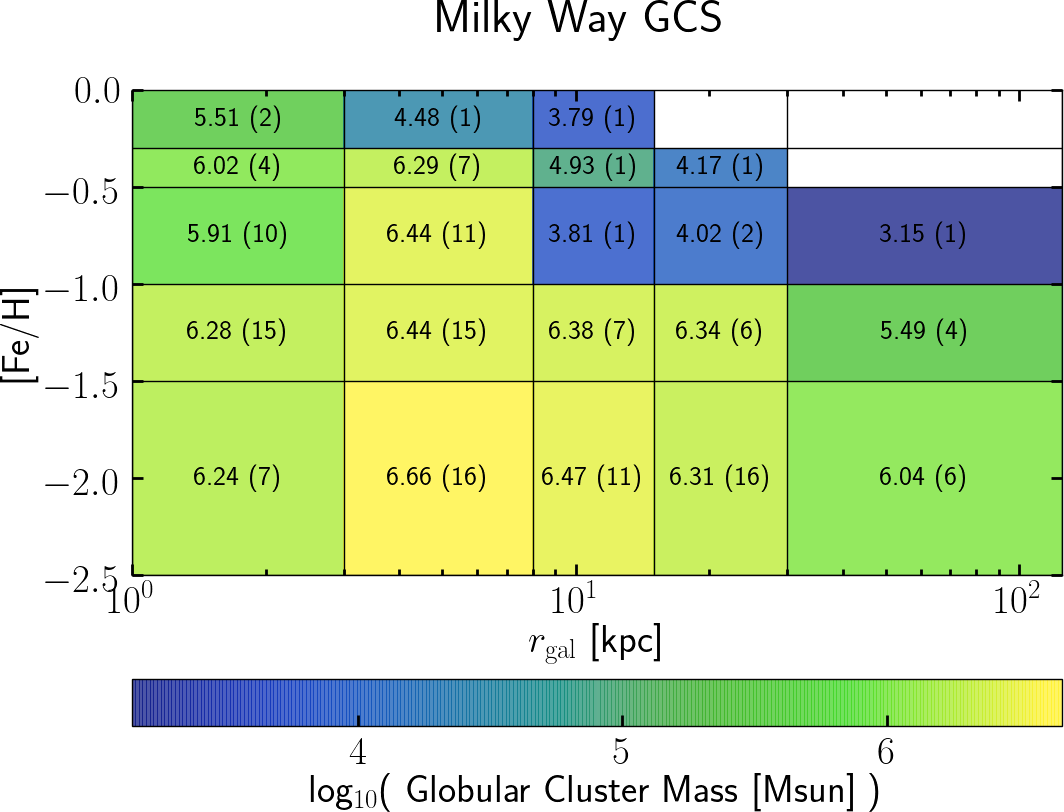}
    \centering \textcolor{lightgray}{\rule[2mm]{\columnwidth}{0.05mm}}
    \includegraphics[width=\columnwidth]
        {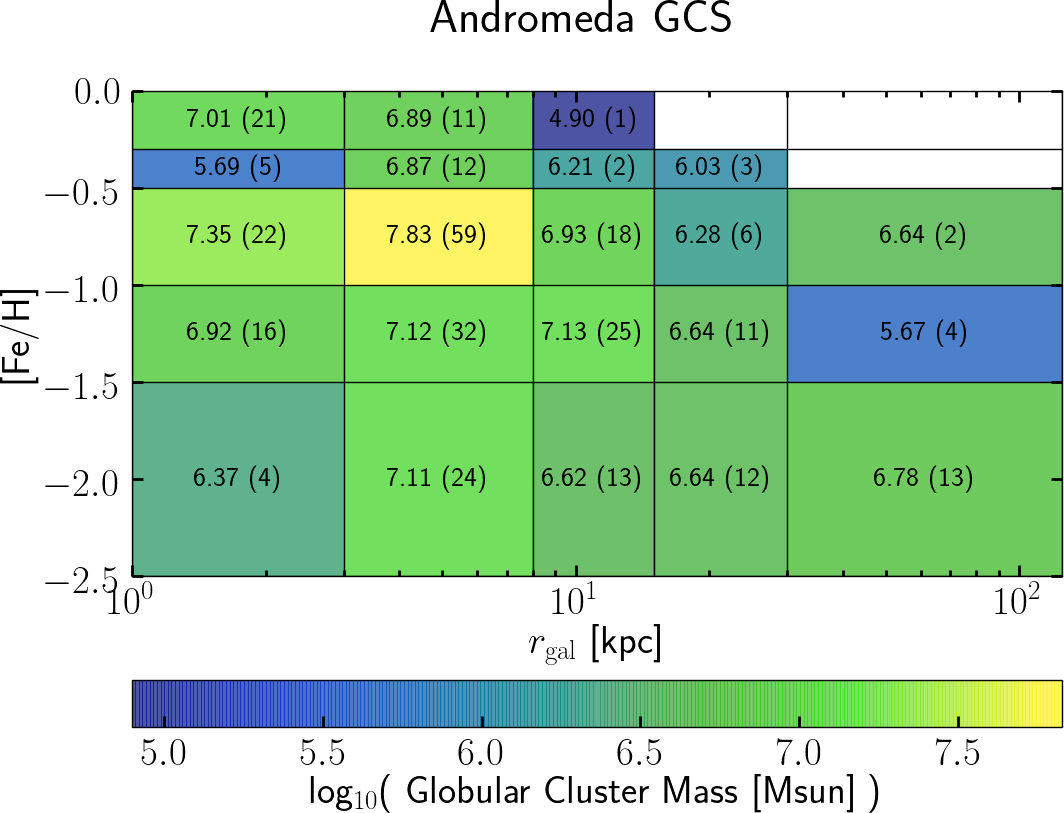}
    \caption{
        \emph{Top}: Mass-weighted $r_{\text{gal}}$-[Fe/H] distribution of
        151 GCs in the MW \citep[data from][2010 ed.]{1996AJ....112.1487H},
        colour-coded by the \textbf{median} (values also shown in each bin). The 
        numbers in parenthesis show how many star particles fall within the bin.
        \emph{Bottom}: Same for M31, showing the 366 GCs in CR16 
        \citep[data from][]{2016ApJ...824...42C}. Radii of M31 are again divided by 
        $r_{\text{vir,M31}}$ and multiplied by $r_{\text{vir,MW}}$. Note that the 
        range of the colourmap differs in both figures,
        \label{fig:observations_FeHRgc}
    }
\end{figure}

We show the two-dimensional mass-weighted metallicity-radius distribution of the
MW and the M31 GCS in the top and bottom panel of Figure~\ref{fig:observations_FeHRgc},
respectively. Note that the number of GCs is different than in 
Figure~\ref{fig:MW-M31-age-Rgc} because here we plot [Fe/H] between -2.5 and 0 
and $r_{\text{gal}}$ between 1 and 250 kpc whereas Figure~\ref{fig:MW-M31-age-Rgc}
shows the full range of metallicities and radii.

The observations indicate very few GCs with high metallicities at large radii (the three bins
in the upper right corner, both for MW and for M31), and relatively few GCs at large
radii in general ($r_{\text{gal}} > 30$~kpc; right column: $11$ GCs or 7.3\% in
the MW and $17$ or 4.6\% in M31). Moreover, M31 hosts more metal-rich ([Fe/H]$ > -1$) 
GCs in each radial bin in comparison to the MW after accounting for the fact that M31
hosts a larger GCS. Finally, given that the marginalized (i.e. the metallicity 
and radial) distributions are not statistically consistent between the two galaxies, 
we find that the two-dimensional distributions are also not consistent. More generally,
the GCS of the MW differs significantly from that of M31, not only in the number of 
clusters but also in their distribution over radius and metallicity. We compare these 
observations to the Auriga simulations in Section~\ref{sec:results_FeHRgc}.

\section{Results}
\label{sec:results}
We define GC candidates in the Auriga simulations as all star particles older
than $10$~Gyr based on the age distribution of the MW GCS (top panel of 
Figure~\ref{fig:MW-M31-age-Rgc}), adopting the definition of a GC from
\citet{2010ARA&A..48..431P} and following the method of 
\citet{2017MNRAS.465.3622R}. However, we caution that this definition, although 
statistically consistent with the estimated ages of all GCs in the MW, is not 
consistent with the tail of younger clusters as stated in Section~\ref{sec:age},
and further discussed in Section~\ref{sec:age_min}.

Throughout our analysis we compare distributions of age-selected GC 
candidates (age $>10$~Gyr) with those of all star particles (i.e.
with no age cut). The mean number of star 
particles in the Auriga L4 simulations that we select as GC candidates is 
$3.1 \cdot 10^5$ (with a mean total mass of $1.15 \cdot 10^{10}$ M$_{\odot}$), 
and the mean number of all star particles is $1.93 \cdot 10^6$ 
($7.3\cdot10^{10}$ M$_{\odot}$). The subset of GC candidates is thus around 
15\% of all the stars in the galaxy when averaged over all Auriga L4 haloes. 
We further split the GC
candidates into an {\it in~situ} subset (defined as GC candidates that are 
bound to the most-massive halo/subhalo in the first snapshot that the particle 
was recorded), and an accreted subpopulation (those that have formed {\it ex~situ} but
are bound to the most-massive halo/subhalo at $z=0$). A GC candidate either formed
{\it in~situ} or it was accreted. We consider the metallicity distribution in 
Section~\ref{sec:results_FeH}, the distribution of galactocentric radii in 
Section~\ref{sec:results_Rgc}, and the combination of the two 
in Section~\ref{sec:results_FeHRgc}.

\subsection{Metallicity distribution}
\label{sec:results_FeH}
\begin{figure}
    \includegraphics[width=\columnwidth]{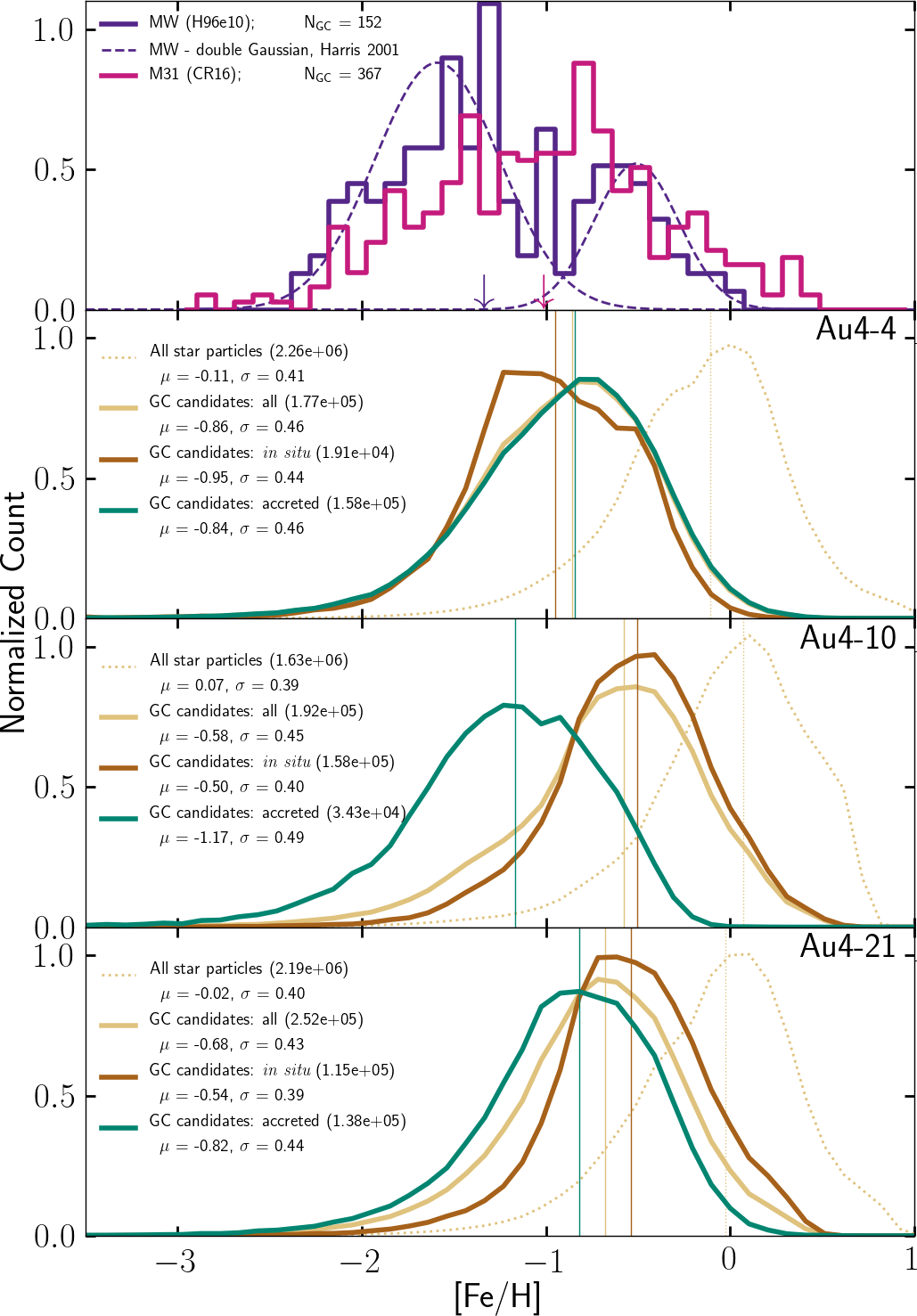}
    \caption{
        \emph{Top}: Metallicity distribution of the MW and M31 GCS. The solid 
        purple (magenta) line shows the GCSs of the MW (M31), and we 
        indicate the mean value with an arrow at the bottom of the x-axis. We
        overplot a double Gaussian for the MW GCS using purple dotted lines, taking
        the mean and dispersion from \citep{Harris2001}.
        \emph{Other panels}: three selected Auriga haloes as indicated in the 
        upper right corner. In each of these three panels we show the GG 
        candidates in beige, and use a beige dotted line for all star particles. 
        We split the GC candidates up into two subpopulations, those that have 
        formed {\it in~situ} (brown), and those that have been accreted (teal).
        The numbers in parenthesis show how many star particles are included 
        in the subset.
        \label{fig:Au4-10and21_FeH}
    }
\end{figure}

We investigate whether the star formation model implemented in Auriga produces 
metallicity distributions consistent with the MW and M31 GCSs, and whether
the simulations generate sufficient total mass at metallicities typical for the
MW and M31 GCSs. To visually inspect the former we show the normalized metallicity 
distribution of three specific Auriga galaxies in Figure~\ref{fig:Au4-10and21_FeH}
in comparison to the MW and M31 [Fe/H] distributions.

We select \mbox{Au4-10}\footnote{The nomenclature is `Au' for Auriga, followed by the
resolution level 4 and halo number 10, indicating which set of initial
conditions was used.}, \mbox{Au4-21}, and \mbox{Au4-4} to show the
distributions of three individual simulation runs that we consider representative
for specific behaviour, and to highlight that different runs give rise to 
different distributions. We show the age-selected GC candidates as well as the 
accreted and {\it in~situ} subsets. 
The top panel of Figure~\ref{fig:Au4-10and21_FeH} shows the MW 
and M31 GCSs, where we overplot a double Gaussian for the MW GCS by 
adopting literature values of the mean $\mu$ and standard deviation $\sigma$ of 
the metal-rich and metal-poor subpopulations \citep{Harris2001}.

We find that the age cut, on average, lowers the mean metallicity $<$[Fe/H]$>$ from 
$0.0$ to $-0.6$. This reflects the enrichment history of the galaxy because the
age cut removes the more metal-rich stellar population which shifts the mean
metallicity towards lower values. Furthermore, accreted GC candidates generally have
lower mean metallicities than the {\it in~situ} subset with differences of roughly
${\sim}0.3$~dex for the majority of the simulation runs. This behaviour can be 
seen in \mbox{Au4-21}, while a slightly larger difference of ${\sim}0.5$~dex is 
seen for \mbox{Au4-10} (as plotted), \mbox{Au4-16}, \mbox{Au4-17}, \mbox{Au4-18} 
and \mbox{Au4-22}. However, this trend is reversed for \mbox{Au4-1} and \mbox{Au4-4} 
for which the {\it in~situ} subset has a lower mean metallicity instead. 
\mbox{Au4-1} is undergoing a major merger at redshift zero and we find a mean
metallicity in this simulation $\mu = -1.51 \, ({-0.74})$ for the {\it in~situ} 
(accreted) GC candidates, although the former consists of only $1019$ particles 
($1.3$\% of all GC 
candidates, and with a total mass of $5 \times 10^{7}$~M\Sun). For \mbox{Au4-4}, 
only ${10.8}$\% of the GC candidates is classified as {\it in~situ} (compared to 
{\it in~situ} fractions of ${40-80}$\% for other haloes). We note that \mbox{Au4-4}
has undergone a major merger around 2~Gyr lookback time, which is probably why
the accreted population is more metal-rich. After inspection of the same figure 
for every one of the $30$~Auriga L4 haloes we find that the simulations produce
(sub)populations of GC candidates that are more metal-rich than the MW and M31 GCSs.
Moreover, none of the simulations produces GC candidates with a bimodal metallicity 
distribution.

\begin{figure}
    \includegraphics[width=\columnwidth]{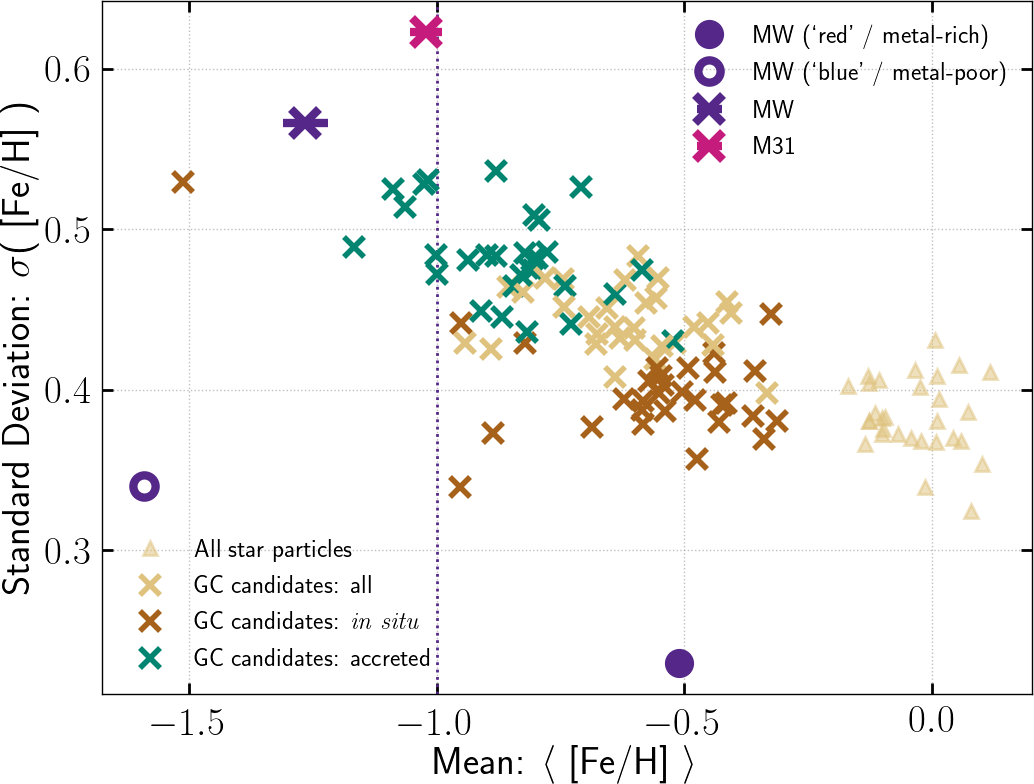}
    \caption{
        First vs. second central moment of the Auriga L4 metallicity distributions.
        Each cross (for a given colour) represents one simulation. The beige (brown)
        [teal] crosses show the values calculated using all (only {\it in~situ})
        [only accreted] GC candidates. Beige triangles indicate that all stars
        were used. The purple (magenta) cross denotes our calculation using all
        MW (M31) observations (which would be appropriate for a unimodal
        distribution). The purple solid (open) dots indicate the literature values
        of a bimodal Gaussian fit to the data \citep[values from][page~38]{1998gcs..book.....A},
        showing the metal-rich (metal-poor) component of the MW (separated at
        [Fe/H] = -1 as indicated by the purple dotted vertical line).
        \label{fig:FeH_mu_sigma}
    }
\end{figure}

We show the mean metallicity and standard deviation of all $30$~Auriga L4 haloes
in Figure~\ref{fig:FeH_mu_sigma} to illustrate the individual behaviour for the 
full set of Auriga galaxies. The beige crosses are to be compared to the purple 
and magenta crosses, which show the mean value of all MW and M31 GCs, respectively. 
In addition, we show the metal-rich population of the MW GCS using a solid dot, 
and the metal-poor population with an open dot. We caution that
these literature values result from individual Gaussian fits to subsets of the 
observational data cut at [Fe/H]~$= -1$. We do not include corresponding data 
points calculated using such an artificial cut for the simulated star particles.
It is interesting that the mean metallicities of the {\it in~situ} GC candidates 
appear roughly consistent with that of the metal-rich population of the MW despite 
the lack of such a metallicity cut, although the simulations show larger dispersions. 
The latter could simply be caused by the hard separation of the MW data into two 
groups, which means the range is smaller and the resulting dispersion lower. With 
regard to the M31 GCS, so far no definitive consensus has been reached
in the literature concerning uni- bi- or trimodality in the [Fe/H] distribution, 
but CR16 argues that the data, after removal of younger objects due to improved 
age classification, hints at three populations separated at [Fe/H]~$=-0.4$ and 
[Fe/H]~$ = -1.5$. None of the Auriga subsets has a mean value that offers much 
hope of reconciling the simulated distribution with the lowest metallicity 
group in the M31 GCS. This is also true for the metal-poor population of the 
MW with a mean value of $-1.6$. The main result of this figure is 
that all Auriga L4 galaxies have metallicity distributions with (much) larger
mean values than observed for the MW and M31 GCS, and that we 
systematically find lower mean metallicities for the accreted GC candidates
than for those that have formed {\it in~situ}.

\begin{figure}
    \includegraphics[width=\columnwidth]{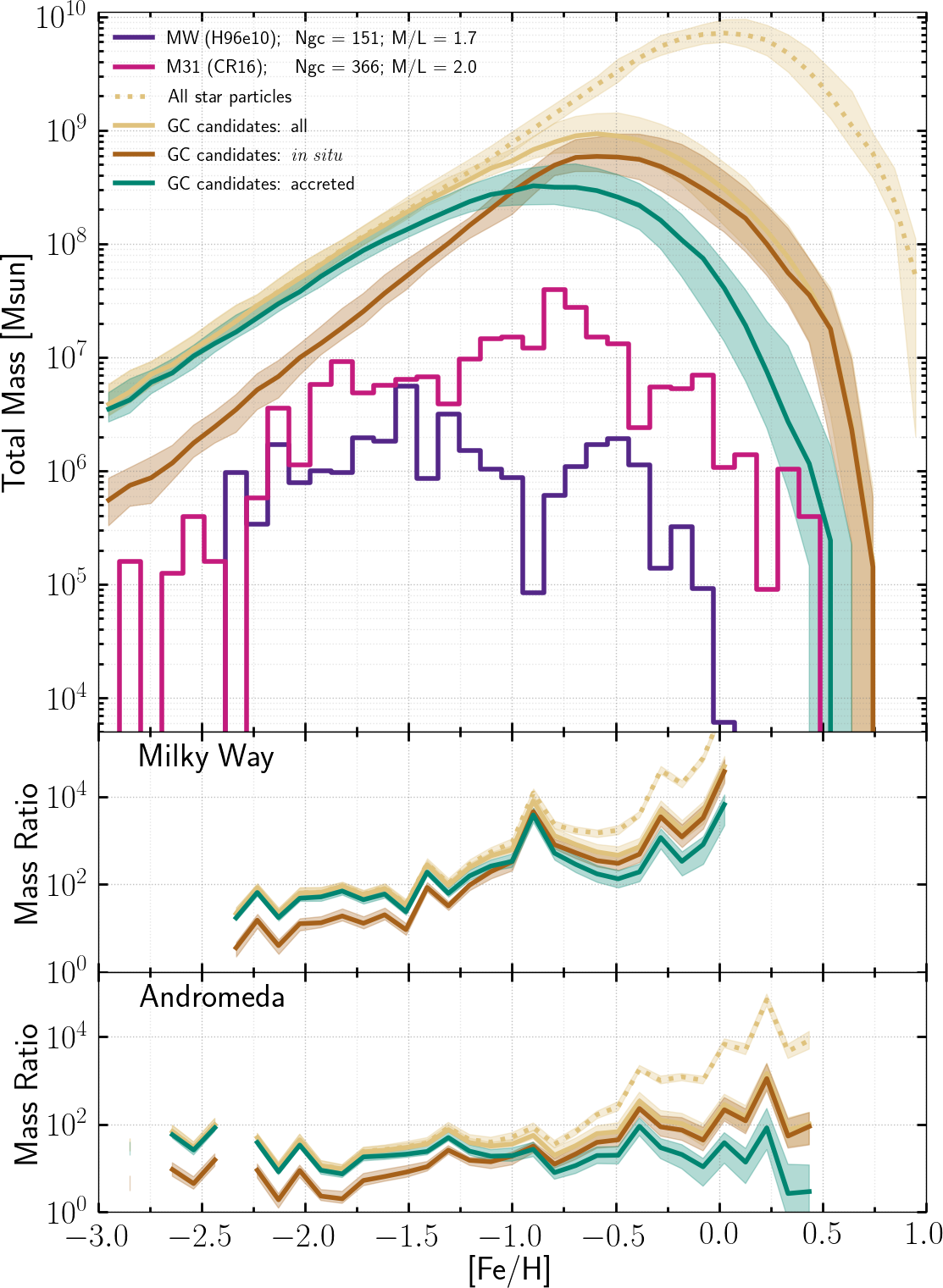}
    \caption{
        Mass-weighted metallicity distribution of star particles in the Auriga
        L4 simulations. We show the median value of all Auriga haloes for all
        stars (beige dotted) and GC candidates (i.e. stars with
        age $>$~10~Gyr; beige solid). The latter subset is further split up
        between stars that formed {\it in~situ} (brown solid), and those that were accreted
        (teal solid). Shaded regions indicate the 25-75 percentile range. The MW (M31)
        GCS is shown in purple (magenta). The middle (bottom) panel shows the
        ratio of the simulated mass to the mass in the MW (M31) GCS.
        \label{fig:FeH}
    }
\end{figure}
We now turn to our second goal, which is to see whether sufficient total mass in 
GC candidates is produced by the Auriga simulations. To answer this question we plot
a mass-weighted metallicity distribution in Figure~\ref{fig:FeH}. We show the 
median over all the $30$~Auriga L4 haloes and the 25-75 percentile range as shaded regions, 
which shows the scatter between runs that have different initial 
conditions, thus, have varying merger histories. We chose to aggregate the data 
to indicate general trends that we find when the GC candidates are split up into
the accreted and {\it in situ} subsets, rather than selecting typical examples 
of individual (simulated) galaxies. Once again we notice that the peak metallicity 
shifts down from [Fe/H]~=~$0$ to [Fe/H]~=~$-0.6$ for GC candidates compared to all stars, 
and we find that the mass at the peak lowers by roughly one dex. The mass budget 
of the GC candidates is dominated by the {\it in~situ} subpopulation above
[Fe/H]~=~$-1$, and by the accreted subset below this value. We show the MW GCS 
in purple and that of M31 in magenta, and notice that the difference between the 
MW and M31 distributions is substantially larger than the scatter between different 
Auriga galaxies, particularly around [Fe/H]~$=-1$. In addition, M31 does host 
(a handful of) GCs with [Fe/H]~$< -2.5$ as well as GCs with [Fe/H]~$ > 0$ while 
the MW does not.

We show the ratio of the  simulated to the observed profiles in the middle and bottom
panels. This mass excess can be thought of the `mass budget' that the Auriga GC 
candidates can `afford to lose' (due to a combination of smaller than unity bound
cluster formation efficiencies combined with a Hubble time of dynamical evolution),
while still producing sufficient mass at the right metallicities. In particular,
the combined efficiency would have to increase with decreasing 
metallicity for Auriga GC candidates to produce a population of GC candidates 
that is consistent with the MW. For the GC candidates in M31 we find a constant 
mass ratio up to [Fe/H]=$-0.9$, above which the simulations produce a higher mass ratio 
with increasing metallicity. If dynamical evolution is not expected to more 
efficiently disrupt GCs of higher metallicity, then we would find that the 
efficiency to form bound star clusters would have to decrease with increasing 
metallicity. This trend could be consistent with the cluster formation efficiency 
model of \citet{2012MNRAS.426.3008K} as lower metallicity star particles formed 
at earlier times when the bound cluster formation efficiencies are higher due
to higher pressure birth environments.

\subsection{Radial distribution}
\label{sec:results_Rgc}

\begin{figure}
    \includegraphics[width=\columnwidth]{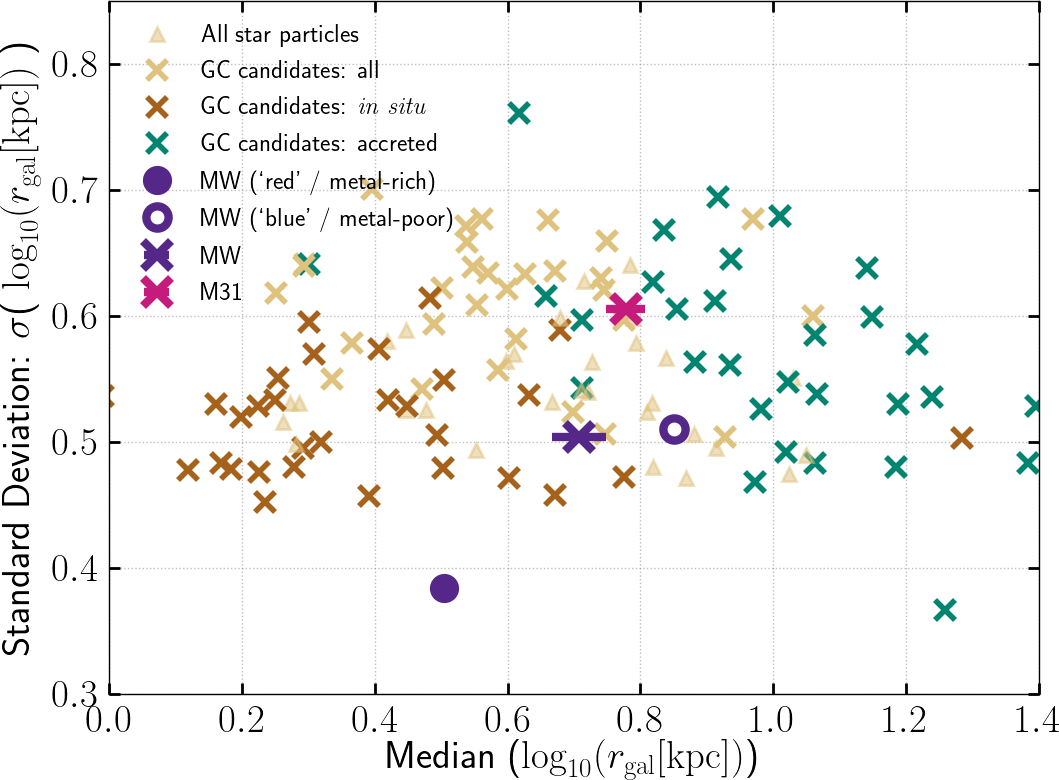}
    \caption{
        Median and standard deviation of the logarithm of radial distribution 
        of star particles in each of the $30$~Auriga L4 haloes compared to
        the MW (M31) GCS shown in purple (magenta). The Auriga (M31) radii are 
        divided by $r_{\text{vir,Au}}$ ($r_{\text{vir,M31}}$) and multiplied 
        by $r_{\text{vir,MW}}$ to compensate for different intrinsic sizes of
        the galaxies.
        \label{fig:Rgc_median_sigma}
    }
\end{figure}
We repeat the analysis of the previous subsection for the distribution of
galactocentric radius instead of metallicity. We divide the Auriga radii by
$r_{\text{vir,Au}}$ and multiply by $r_{\text{vir,MW}}$ to compensate for the different
intrinsic sizes of the Auriga galaxies. We look for general trends present
in all Auriga L4 galaxies. Figure~\ref{fig:Rgc_median_sigma} shows the median and 
standard deviation of $\log_{10}(r_{\text{gal}})$ of star particles in all Auriga
L4 simulations, respectively. The purple and magenta cross show the median
of the MW and M31 GCs, which are to be compared to the beige crosses. The solid
dot now shows our calculation of the median value of the `red' metal-rich population,
and the open dot that of the `blue` metal-poor population. We split the radii 
up into the metal-rich/metal-poor groups by taking a metallicity cut at 
[Fe/H]~=~-1 as is done in the literature.

We notice that the {\it in~situ} GC candidates are more centrally distributed, 
whereas the accreted subsets have a larger radial extent. This 
is not surprising because the classification of {\it in~situ} requires star particles 
to have formed within the virial radius, thus they could naturally be expected 
to end up at small galactocentric radii. Accreted star particles, on the other 
hand, have formed in another (sub)halo beyond the virial radius, thus, they would 
first have to migrate inwards in order to populate the innermost radii. Moreover,
we find that the simulations have a wider dispersion in $\log(r_{\text{gal}})$ than
the MW GCS, while the dispersion of M31 seems to lie within the range of dispersions
found in the Auriga galaxies. Furthermore, the median of the MW GCS lies roughly 
within the range of values produced by the Auriga simulations, whereas the median
of M31 is somewhat larger and slightly closer to typical median values of the 
accreted GC candidates in the Auriga simulations. The larger radial extent 
of the M31 GCS is generally believed to hint at a more accretion-dominated origin 
of the GCS, and may reflect a richer accretion history of M31 in comparison 
to the MW.

\begin{figure}
    \includegraphics[width=\columnwidth]{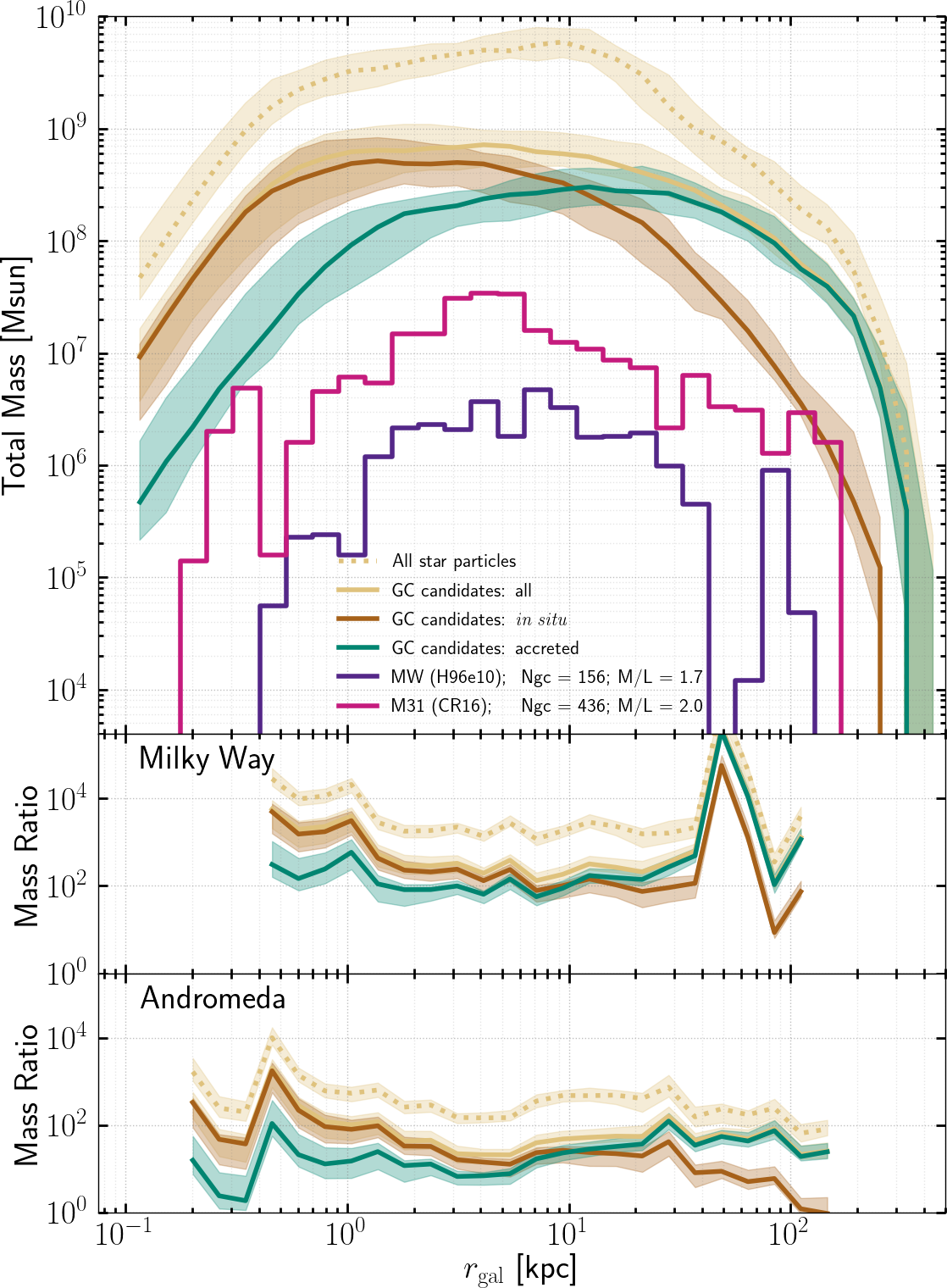}
    \caption{
        Mass-weighted radial distribution of star particles in the Auriga
        L4 simulations, where the Auriga (M31) radii are again divided by 
        $r_{\text{vir,Au}}$ ($r_{\text{vir,M31}}$) and multiplied by 
        $r_{\text{vir,MW}}$. We show the median value of all Auriga haloes for all
        stars (beige dotted) and GC candidates (i.e. stars with
        age $>$~10~Gyr; beige solid). The latter subset is further split up
        between stars that formed {\it in~situ} (brown solid), and those that 
        were accreted (teal solid). Shaded regions indicate the 25-75 percentile
        range. The MW (M31) GCS is shown in purple (magenta). The middle 
        (bottom) panel shows the ratio of the simulated mass to the mass in the 
        MW (M31) GCS.
        \label{fig:Rgc}
    }
\end{figure}
Figure~\ref{fig:Rgc} shows the mass-weighted radial distribution of the
Auriga L4 haloes. We notice a subtle peak around $10$~kpc for all star
particles that is not present for the GC candidates, indicating that the
stellar disc is no longer present when applying the latter selection criterion.
Furthermore, we find that the dominant contribution to the total mass in GC
candidates changes from those formed {\it in~situ} to the accreted population around
$10$~kpc. Again we show the mass ratio of the simulations compared to the MW 
and M31 GCS. We find a decreasing mass ratio with increasing radius
in the range $0.2$ to ${\sim}5$ kpc, followed by an increase attributed
to the accreted subpopulation. For the MW we notice that significantly
fewer GCs are found beyond $40$~kpc than for M31, and that accreted GC candidates
contribute mostly to the Auriga GCS at these radii. We further investigate a 
breakdown of the total mass in Auriga GC candidates into bins of both metallicity 
and radius in the following section.

\subsection{Total mass in metallicity-radial space}
\label{sec:results_FeHRgc}

\begin{figure}
    \includegraphics[width=0.49\textwidth]
        {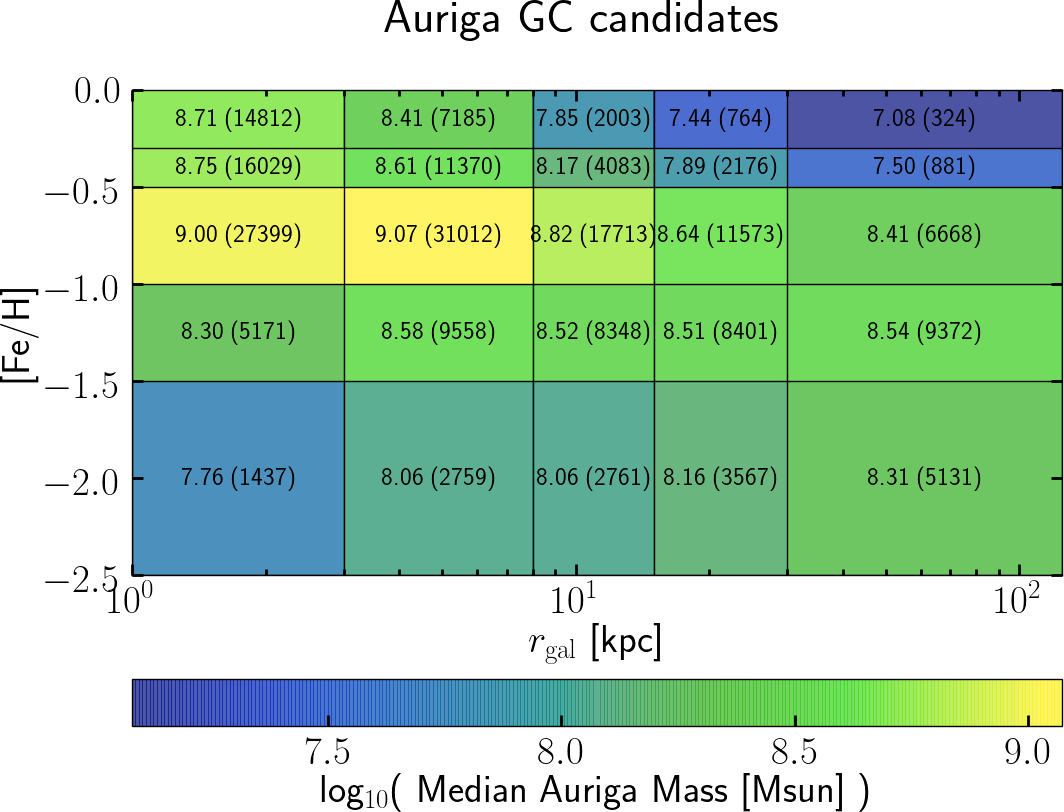}
    \caption{
        Mass-weighted [Fe/H]-$r_{\text{gal}}$ distribution of all $30$~Auriga
        L4 haloes. Here we consider only the GC candidates and colour-code by the 
        median mass (values also shown in each bin). The numbers in parenthesis 
        show how many star particles fall within the bin. Note that the range of 
        the colourmap differs from both panels in Figure~\ref{fig:observations_FeHRgc} 
        (for improved contrast within the plot).
        \label{fig:Au-FeHRgc}
    }
\end{figure}

We investigate whether the Auriga simulations still produce sufficient mass when
the GC candidates are two-dimensionally binned in [Fe/H] and $r_\text{gal}$. First
we sum the total mass of GC candidates in each bin (for an individual Auriga 
simulation), then we calculate the median total mass in each bin over all 
$30$~Auriga L4 haloes. The result is shown in Figure~\ref{fig:Au-FeHRgc}, which 
can be compared to the observed total GC mass shown in 
Figure~\ref{fig:observations_FeHRgc}. We notice that the simulations produce
relatively little mass in metal-poor GC candidates at small radii and in metal-rich
GC candidates at large radii. However, there is still few times $10^7$ M$_{\odot}$
produced by the simulations in the uppermost 3 bins where no observed GCs are found.

\begin{figure}
    \includegraphics[width=0.49\textwidth]
        {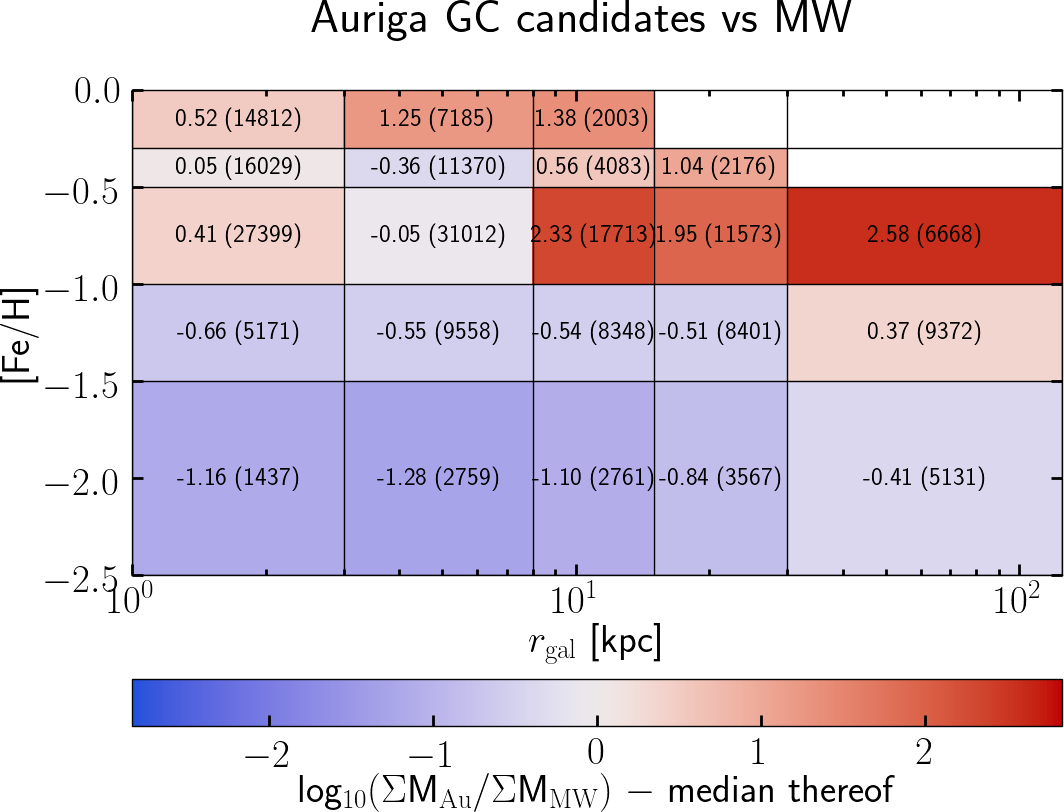}
    \centering \textcolor{lightgray}{\rule[2mm]{\columnwidth}{0.05mm}}
    \includegraphics[width=0.49\textwidth]
        {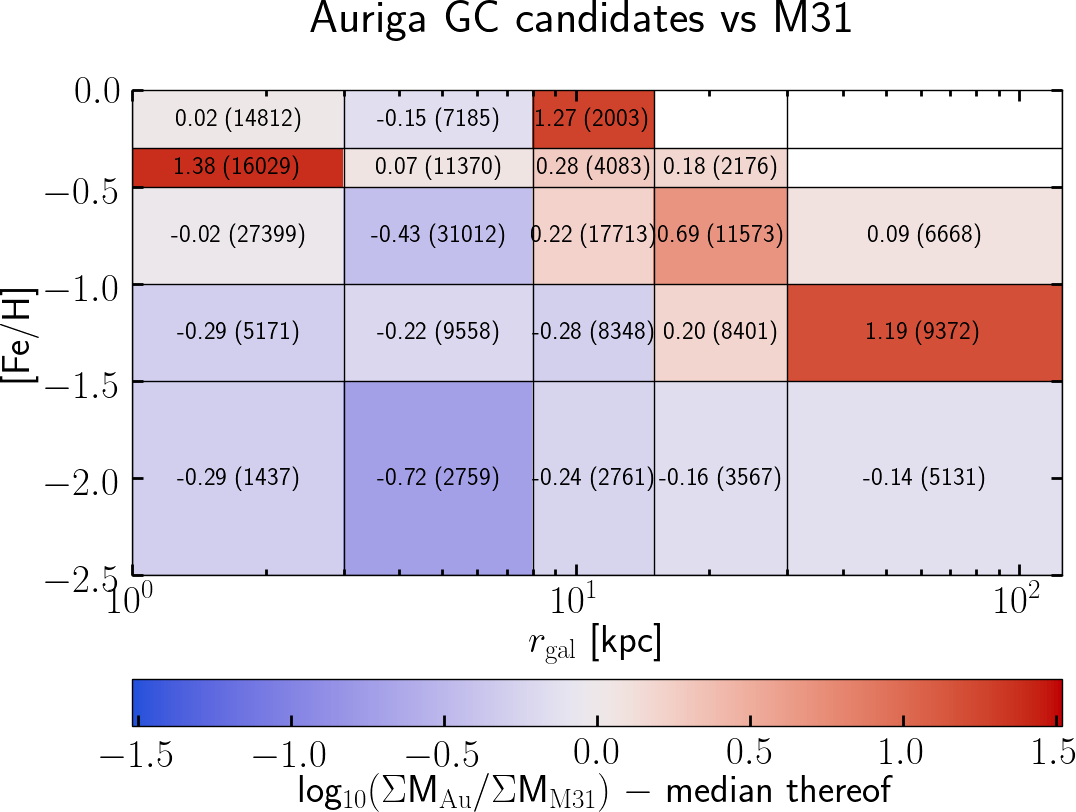}
    \caption{
        The top (bottom) panel shows the logarithm of the ratio of simulated mass
        to mass in the MW (M31) GCS, i.e. the logarithm of the mass ratio, minus
        the median value of these 25 bins. The values are 2.68 and 1.69 dex for 
        the MW and M31, respectively. Red (blue) indicates a higher (lower) overall 
        mass excess. The number in each bin shows the colour-coded value that is 
        plotted, and the (median) star particle count is given in parenthesis. 
        Note that the three bins in the upper right corner are white because the 
        observations have zero mass there.
        \label{fig:Au-FeHRgc-ratio}
    }
\end{figure}

To ease this comparison between the simulations and observations we take the ratio
of the total mass of GC candidates in Auriga to the total GC mass in the observations.
We use a diverging colour map centered around the overall median of the 
mass ratio (i.e. we subtract the median of these $25$ bins from the value 
in each bin). We show this in the top and bottom panel of Figure~\ref{fig:Au-FeHRgc-ratio}
for the MW and M31, respectively. Note that the 3 bins in the top right are empty (white)
because the mass ratio cannot be calculated due to zero mass in the observations.
If the product of the bound cluster formation efficiency and consecutive GC 
disruption over a Hubble time would be constant, then the value in each bin would
be zero. The red and blue bins indicate deviations from such a scenario and
indicate a higher and lower than average mass excess in the simulations. The median 
value used to shift the mass ratio is $2.68$ and $1.69$ dex for the MW and M31, 
respectively. This means that the median `mass budget' that the simulations `can 
afford to lose' is roughly a factor $480$ and $50$ for the MW and M31, respectively. 
The lowest mass surplus for both galaxies is the bin with metallicities in the range
[Fe/H]~$\in [-2.5, -1.5]$ and galactocentric radius $r_{\text{gal}}\in [3,8]$~kpc,
for which the simulations produce 
a factor $10^{2.68}/10^{1.33} \approx 22$ and $10^{1.69}/10^{0.75} \approx 9$ 
more mass in GC candidates than is observed in the MW and M31 respectively.
The highest factor for the MW is $2.1 \times 10^{5}$, and $1.1 \times 10^{3}$
for M31.

We now focus on four quadrants of this diagram, split at [Fe/H]~=~-1 (metal-poor 
or metal-rich) and $r_\text{gal}$~=~8~kpc (inside or outside the Solar radius).
The majority of the mass in Auriga GC candidates is produced in metal-rich star
particles inside the Solar radius. The total mass is dominated by the {\it in situ} 
component, for [Fe/H] as well as for $r_\text{gal}$ as can be seen in Figure~\ref{fig:FeH}
and Figure~\ref{fig:Rgc}. The total mass excess in 
this region increases with decreasing metallicity for the MW and M31, and there
is a weak trend of decreasing over-production with increasing radius for the
{\it in situ} subsets but no trend with radius for the accreted GC candidates.
As for the metal-rich GC candidates outside the Solar radius, we find that the
mass excess with respect to the median is largest in this corner for the MW and
M31. The main reason for this trend are the lower GC counts in the observations,
whereas the Auriga simulations efficiently form metal-rich GC candidates and
have no problem to produce them at large radii. The opposite is true for the
metal-poor GC candidates inside the Solar radius, where the mass surplus with
respect to the median is smallest for both galaxies. However, the disruption rates
are expected to be particularly high for GCs in the inner regions of galaxies
which must have orbits with pericentres close to the galactic bulge or bar. In addition, 
disk crossings in this region will cause spikes in the tidal tensor due to steep
density gradients that could be highly disruptive to the GCs. The bins with the 
lowest `mass budget' are in this region (a factor 9 for the MW and 22 for M31).
The bound cluster formation efficiency has to exceed 10\%, and none of the GC 
candidates in this region can experience significant mass loss in order for Auriga
to produce sufficient mass in this quadrant. Finally, the metal-poor GC candidates
outside the Solar radius have lower than average mass ratios, which is not 
surprising given that the Auriga simulations overall underproduce low-metallicity
stars particles. 

\section{Discussion}
\label{sec:discussion}
\subsection{Metallicity distribution}
\label{sec:discussion_FeH}
GCs are ${\sim}0.5$ dex more metal-poor than spheroid stars observed 
at the same radius for almost all galaxies \citep{1991ARA&A..29..543H}. Our 
selection function (age cut) does lower the mean metallicity of GC candidates by 
$0.5$ dex with respect to all star particles, but we still find that the GC 
candidates in the Auriga simulations are more metal-rich than the MW and M31 GCSs. 
Although the (old) star particles represent single stellar populations with a mass 
consistent with that of GCs, they are in fact statistical tracers 
of the stellar population of the galaxy as a whole. Therefore only a (small) 
fraction of the star particles may represent plausible formation sites of GCs, 
whereas the majority represents (halo) field stars - the disk component effectively 
falls outside our selection of star particles due to the age cut. It is implausible 
that all halo stars come from disrupted GCs because the total mass in GCSs 
at redshift zero is generally a few orders of magnitude lower than the total 
mass of stellar haloes of galaxies. If stellar haloes solely consist of disrupted
GCs, then the GC disruption rate would have to be considerably higher than expected
from theoretical models. Moreover, disrupted dwarf galaxies are more likely to 
contribute to the build up of the stellar halo than GCs \citep[e.g.][]{2015MNRAS.448L..77D, 
2019ApJ...887..237C}. Overall, our set of GC candidates most likely contains
an overabundance of regular stars. It could be that the resulting metallicity
distribution is then biased towards higher values than observed for real 
GCs. An alternative explanation could overmixing of metals at early times.

In general, the colour (metallicity) distribution of most GCSs appears to be bimodal 
with typical separating value [Fe/H]$\sim -1$ \citep{1985ApJ...293..424Z,
1999AJ....118.1526G,2001AJ....121.2974L,2006ApJ...639...95P}. Indeed, 
\citet{1998gcs..book.....A} and \citet{Harris2001} find that the MW GCS has a 
bimodal [Fe/H] distribution. Observations of GCs in M31, however, may be best 
split into three distinct metallicity groups \citep{2016ApJ...824...42C}. The 
numerical simulation of \citet{2017MNRAS.465.3622R} does yield a bimodal 
metallicity distribution where the metal-poor population is dominated by accreted 
star particles and the metal-rich population by {\it in~situ} stars. 

However, we find that none of the Auriga simulations produces a bimodal 
metallicity distribution for age-selected GC candidates. Interestingly, the 
cross-over point above (below) which the mass-weighted metallicity distribution
of GC candidates is dominated by those that have formed {\it in~situ} (were
accreted) does coincide with the separation between the metal-rich and metal-poor
populations of GCSs at [Fe/H]~=~$-1$, as for the MW. Moreover, we do find 
that the mean metallicity shifts when we split the GC candidates according 
to birth location (accreted or {\it in~situ}). In particular, the mean values of 
the {\it in~situ} GC candidates are roughly consistent with the metal-rich MW GCs.
On the other hand, we also notice that most of the simulations have a similar 
mean value for all GC candidates, and the simulated {\it in~situ} GC candidates 
have larger dispersions than is claimed for metal-rich GCs in the MW. Moreover, the 
offset between the means of the metal-rich and metal-poor populations in the MW is
$1$ dex, a factor $2-3$ larger than the offset that we find between the {\it in~situ}
and accreted populations. However, we caution that \citet{2019MNRAS.486.3134K} suggests 
to reserve the `{\it ex~situ}' classification for accretion after $z=2$, when
the central galaxy has formed and accretion unambigously contributes to the
radially extended halo GC population. In fact, our classification might not hold for Au4-1.

Finally, we note that \citet{2006ARA&A..44..193B} compares the number of metal-poor 
GCs to the stellar halo mass and find\footnote{The specific frequency $T$ is 
the number of GCs per $10^9$ M\Sun \, of galaxy stellar mass.} $T^n_{\text{blue}} \sim 100$, 
while the number of metal-rich GCs compared to the bulge mass yields 
$T^n_{\text{red}} \sim 5$, and therefore conclude that the formation efficiency
of metal-poor GCs is 20 times higher than the metal-rich GCs with respect to 
field stars. We find that a gradual increase in formation efficiency of GCs
with respect to field stars would be required with decreasing metallicity for
the GC candidates in the Auriga simulations to yield sufficient total mass to
be consistent with the MW or M31 GCS. In addition, the metal-poor GC candidates
would have to experience low mass loss rates, especially those within the Solar
radius.

\subsection{Radial distribution}
\label{sec:discussion_Rgc}
The M31 GCS has a factor 2-3 more GCs at large radii compared to the MW GCS,
which may indicate that M31 has a richer merger history than the MW 
\citep{2016ApJ...824...42C}. In the Auriga simulations we find that the GC 
candidates at radii larger than ${\sim}20$~kpc are indeed dominated by accreted
star particles. However, we select star particles that are bound to the main 
halo and main subhalo, which means that we include particles up to the virial 
radius $r_{200}$. The Auriga simulations have no problem populating the stellar 
halo up to the virial radius, even with our additional age cut. On the other hand, 
the MW and M31 have fewer GCs at large radii. We expect tidal disruption to be
less efficient at larger radii, thus, the formation efficiencies of the accreted 
GC candidates would have to be lower Alternatively, the fraction of old star 
particles in our subset of GC candidates may contain considerably more star 
particles that represent stars in the old stellar halo that have been accreted
during hierarchical buildup of the galaxy. Our classification of GC candidates 
could be improved for the accreted subset in particular. For example, we could 
select only a subset of the accreted star particles as GC candidates where the 
number number of GCs would be determined by the satellite halo mass, given the 
$M_{vir}-N_{\text{gal}}$ correlation found by \citet{2019arXiv190100900B}. However,
we did not include such an analysis in this study.

\subsection{Lower age cut in the GC selection function}
\label{sec:age_min}
Our method is sensitive to the definition that we adopt to classify GC candidates
\citep[e.g.][]{2018RSPSA.47470616F}. Therefore we explore how our results change if we 
include all star particles older than 8 Gyr, and if we further lower the age cut 
to 6 Gyr. The average fraction of GC candidates that we select from the all the
star particles increases to 35\% and 55\% for an age cut of 8 and 6 Gyr, respectively.

We find that including more recently formed star particles in our subset 
of GC candidates shifts the mean metallicity to higher values. Moreover, the
total mass in metal-rich ([Fe/H] > -1) GC candidates increases for a lower age 
threshold and moves closer to the dotted line that shows the distribution of
all star particles in Figure~\ref{fig:FeH}. However, 
we notice that the Auriga simulations, on average, do not form new stars with 
a metallicity below [Fe/H]~=~-1 more recent than 6 Gyr ago: the behaviour of 
Figure~\ref{fig:FeH} does not change for metallicities below [Fe/H]~=~-1 when 
we change the age cut to 8 or 6 Gyr, and the slope of the masss ratio for 
metallicities above [Fe/H]~=~-1 is slightly steeper. In addition, we still find
that the accreted GC candidates are more metal-poor than those that have formed 
{\it in~situ}. 

The qualitative behaviour of the median distribution of galactocentric radii 
(Figure~\ref{fig:Rgc}) is unaffected by lowering the age cut other than including
more star particles (more mass) which leads to a higher normalisation and
a higher mass ratio with respect to the observed GC systems of the MW and M31.
The qualitative behaviour presented in Figure~\ref{fig:Au-FeHRgc-ratio} does not 
change when the age cut is lowered to 8 or 6 Gyr. 
Overall, reducing the age cut worsens the agreement between the observed GCSs 
and the candidate GCs in the Auriga galaxies.

\section{Summary and conclusions}
\label{sec:conclusions}

We investigate statistics of age-selected ($>$~10~Gyr) GC candidates in the 
Auriga simulations and compare the simulations to the MW and M31 GCSs to test
whether the star formation model implemented for the Auriga simulations could give
rise to metallicity and radial distributions that are consistent with these two
observed GC systems. Based on our analysis we draw the following conclusions.

\begin{itemize}
    \item The star formation model implemented in the Auriga simulations produces
    metallicity distributions that are more metal-rich than the MW and M31
    GCSs. For most of the Auriga simulations we find that the subset of accreted 
    GC candidates has a lower mean metallicity than the {\it in situ} subpopulation.
    However, the accreted subsets are still more metal-rich than the MW and M31 GCSs.
    Moreover, none of the simulated subsets has a mean metallicity as low as the mean
    of the `blue' metal-poor component of the MW GCS.

    \item We find that the difference between the metallicity distributions of 
    the MW and M31 GCSs is big compared to the scatter in the Auriga GC candidates 
    of different Auriga runs. The radial distributions of the MW and M31 GCSs are
    somewhat more similar, but they are not consistent with being drawn from the
    same distribution. The scatter between the radial distributions of different
    Auriga simulations is smaller than the difference between the MW and M31 GCSs.

    \item GC candidates in the Auriga simulations may be found out to $r_{200}$,
    given that our selection function of star particles selects all stars bound
    to the main subhalo of the halo. The total mass of GC candidates is 
    dominated by accreted star particles at radii beyond $20$~kpc. The GCs in 
    the MW and M31, on the other hand, have a much smaller radial extent.

    \item Tidal forces are larger in the inner regions of galaxies than in the
    outer halo. Therefore, we expect dynamical evolution to more strongly affect 
    GC candidates at small radii than at large radii. However,
    we find that the Auriga simulations produce more mass in GC candidates with
    high metallicities at large radii than with low metallicities at small radii.
    The cluster formation efficiency would have to increase with decreasing
    metallicity for GC candidates in the Auriga simulations to be consistent
    with the MW GCS. This trend of over-production of old star particles that 
    are metal-rich and at large radii compared to observed GCs is less clear for 
    the M31 GCS.

    \item
    Overall, we conclude that the metallicity and radial distribution of age-selected
    star particles in the Auriga simulations are not consistent with globular 
    cluster system of the Milky Way or with that of the Andromeda galaxy.

\end{itemize}

\section*{Acknowledgements}
We thank the anonymous referee for a thoughtful and elaborate report.
TLRH acknowledges support from the International Max-Planck Research School (IMPRS)
on Astrophysics. TLRH thanks the Max Planck Computing and Data Facility for
maintaining the Freya compute cluster, where the simulations were post-processed.
FM is supported by the program "Rita Levi Montalcini" of the Italian MIUR. FAG 
acknowledges financial support from CONICYT through the project FONDECYT Regular 
Nr. 1181264, and funding from the Max Planck Society through a Partner Group grant.
This work has been supported by the European Research Council under ERC-StG grant
EXAGAL- 308037. Part of the simulations of this paper used the SuperMUC system 
at the Leibniz Computing Centre, Garching, under the project PR85JE of the Gauss 
Centre for Supercomputing. This work used the DiRAC Data Centric system at Durham
University, operated by the Institute for Computational Cosmology on behalf of
the STFC DiRAC HPC Facility `www.dirac.ac.uk'. This equipment was funded by BIS 
National E-infrastructure capital grant ST/K00042X/1, STFC capital grant 
ST/H008519/1 and STFC DiRAC Operations grant ST/K003267/1 and Durham University.
DiRAC is part of the UK National E-Infrastructure.

The analysis in this work was performed using the Python \citep{python}
programming language, the IPython \citep{2007CSE.....9c..21P} environment, 
the NumPy \citep{2011CSE....13b..22V}, SciPy \citep{scipy}, and Astropy 
\citep{2013A&A...558A..33A} packages. Plots were created with Matplotlib 
\citep{2007CSE.....9...90H}, using the perceptually uniform colour maps
from \citet{2015arXiv150903700K} for two-dimensional plots. This research 
has made use of NASA's Astrophysics Data System Bibliographic Services.




\bibliographystyle{mnras}
\bibliography{AurigaGCS}




\bsp    
\label{lastpage}
\end{document}